\documentclass[a4paper,fleqn]{cas-dc}

\usepackage{placeins}
\usepackage{float}
\usepackage{ragged2e}
\usepackage[numbers]{natbib}
\newcommand{\tabincell}[2]{\begin{tabular}{@{}#1@{}}#2\end{tabular}}
\def\tsc#1{\csdef{#1}{\textsc{\lowercase{#1}}\xspace}}
\tsc{WGM}
\tsc{QE}
\tsc{EP}
\tsc{PMS}
\tsc{BEC}
\tsc{DE}

\begin{document}
\let\WriteBookmarks\relax
\def\floatpagepagefraction{1}
\def\textpagefraction{.001}

\shorttitle{Leveraging social media news}

\shortauthors{CV Radhakrishnan et~al.}

\title [mode = title]{Quantifying sex differences in brain network topology by aggregating nodal centrality rankings}                      
\tnotemark[1,2]

\tnotetext[1]{This document is the results of the research
   project funded by the National Science Foundation.}

\tnotetext[2]{The second title footnote which is a longer text matter
   to fill through the whole text width and overflow into
   another line in the footnotes area of the first page.}

\author[1]{Wenyu Chen}[type=editor,
                        auid=000,bioid=1,
                        role=Researcher,
                        orcid=0000-0003-4299-8764]
\ead{shanyu618@163.com}


\address[1]{College of Computer and Information Science, Southwest University, Chongqing, 400715, P. R. China} 

\author[1]{Ling Zhan}[style=chinese]

\author[1]{Yunsong Luo}[style=chinese]

\author[2,3,4]{Jiang Qiu}[%
   ]
\cormark[1]
\ead{qiuj318@swu.edu.cn}


\address[2]{Key Laboratory of Cognition and Personality (SWU), Ministry of Education, Chongqing, China}
\address[3]{Faculty of Psychology, Southwest University, Chongqing, China}
\address[4]{Southwest University Branch, Collaborative Innovation Center of Assessment Toward Basic Education Quality at Beijing Normal University, Beijing, China}
\author%
[1]
{Tao Jia}
\cormark[2]
\ead{tjia@swu.edu.cn}

\begin{abstract}
Although numerous studies report significant sex differences in functional connectivity, these differences do not sufficient to reveal specific functional disparities among brain regions or the topological differences in brain networks. Meanwhile, individual differences could potentially bias the understanding of these sex differences. To address these challenges, we propose a consensus rank-based method to quantify sex differences in four node centrality ranking within the functional brain network. 
This method aggregates individuals' nodal centrality rankings into a consensus or "average" ranking, minimizing the impact of outliers and enhancing the robustness of the findings. 
By analyzing resting-state functional MRI data from 1,948 healthy young adults (aged 18-27 years, 1,163 females), we find significant sex differences in the topology of functional brain network, primarily attributed to biological sex rather than individual differences.
Particularly, sex accounts for approximately 10\% of the differences in nodal centrality consensus rankings. Using a rank difference index (RDI), we identify eight critical brain regions with the greatest rank differences, including the insula, supramarginal gyrus, and dorsolateral superior frontal gyrus. females show higher rankings in regions with stronger intra-system connections, whereas males dominate in areas with stronger inter-system connections.
Our findings enhance our understanding of sex-specific characteristics in functional brain networks. 
Moreover, our approach may offer novel insights into targeted population studies, including those involving healthy individuals and patients with brain injuries. 


\end{abstract}


\begin{highlights}
\item A novel consensus rank-based method for quantifying sex differences in functional brain network topology is proposed.
\item The observed sex differences in functional brain networks are attributed to biological sex rather than individual differences.
\item Sex accounts for around 10\% of differences in nodal centrality consensus rankings.
\item We identify eight hub regions exhibiting the most significant rank differences between females and males. 
\end{highlights}


\maketitle

\section{Introduction}
Sex differences have garnered considerable attention in physiological research due to their implications for cognitive abilities and behavior \citep{lewin2002sex, levine2016sex,satterthwaite2015linked,hyde2016sex,jancke2018sex}. Understanding the interplay between biological sex and the brain is pivotal for shedding light on sex-related variations in cognitive tasks such as language processing, social cognition, memory, and emotion recognition. Resting-state functional magnetic resonance imaging (rs-fMRI) \citep{biswal1995functional,fox2007spontaneous} has facilitated the exploration of sex differences in the neural mechanisms underlying cognitive processing. Recent studies have revealed significant sex differences in brain anatomy at rest, including variations in gray matter volume, white matter volume, and cortical thickness, which may contribute to differences in cognitive task performance observed between males and females \citep{ritchie2018sex, sanchis2019sex, wang2019hemisphere, wierenga2018key, jancke2018sex}. 
Although anatomical differences have been observed, these structural variations alone may not fully explain how sex influences brain function \citep{joel2020complex}. 
Hence, it is essential to further explore the underlying mechanisms that contribute to sex differences in brain function. 

Currently, the predominant approach to investigating sex differences in resting-state brain function involves examining differences in functional connectivity (i.e., the connection between two regions of interest) \citep{satterthwaite2015linked, rubin2017sex, zhang2016sex, ritchie2018sex, wang2019hemisphere, iccer2020functional, stumme2020functional}. These investigations generally agree that females tend to exhibit stronger connections within the default mode network (DMN) and the visual network (VIN), while males demonstrate stronger connections within the salience network (SN) and between the sensorimotor networks and other functional networks. However, employing connection matrix comparisons in brain network analyses presents several limitations. We can only identify different functional connectivity by comparing two matrices, but the differences in brain regions are usually of greater interest. Most importantly, even when distinct connections are identified, they may not fully reflect the topological changes in the functional brain network. For instance, a connection may hold the same weight in two brain networks; however, in one network, it may act as a bridge connecting two isolated components with high betweenness, while in the other network, it may function as one of many regular connections without significant roles. Focusing solely on the connection's weight may lead to erroneous conclusions, as it overlooks the different topological roles this connection plays in the two networks.

Due to the aforementioned limitations, several studies have begun employing different node centrality measures to quantify the functional importance of brain regions \citep{van2013network,power2013evidence,rodrigues2019network}. These measures involves initially constructing an individual's network from rs-fMRI data and then identifying central nodes within this network using established network science theories. Subsequently, statistical comparisons are conducted to discern differences between the two groups \citep{yan2011sex,zuo2012network,tian2011hemisphere,crossley2014hubs,van2019cross,mheich2020brain}. While differences in centrality scores across two groups can be ascertained \citep{tian2011hemisphere, zuo2012network}, it is essential to note that centrality scores reflect the relative importance of a brain region within a specific network and may not be directly comparable across individuals. For example, a brain region might serve as a hub node in one network but merely as a non-hub node in another. 
Hence, we employ the centrality ranking method for brain regions. It facilitates a uniform comparison of the importance of brain regions across different individuals, overcoming the variability inherent in individual network structures.

In addition, while statistical comparisons such as two-sample t-tests or analysis of variance (ANOVA) can be conducted to compare the centrality rankings of the same brain regions between two groups, averaging these rankings may obscure the consistent importance of brain regions at the group level, especially if the rankings have skewed distributions. This approach may contribute to the argument that observed differences between males and females in the brain characteristics are influenced by individual differences \citep{cahill2006sex,kaczkurkin2019sex,cosgrove2007evolving, grabowska2017sex}. Moreover, averaging individual rankings may obscure critical details about the relative importance of different brain regions.

In the present study, we aim to address the above issues using a large rsfMRI dataset of 1,948 healthy young adults (60.3\% female, age range 18-27 years) to investigate sex-related effects on topological structure of functional brain network. 
We focus on four different node centrality indicators, degree centrality, eigenvector centrality, closeness centrality, and betweenness centrality, which reflect the brain network's local and global topology respectively.  
To facilitate group comparisons, we employ a ranking aggregation method to obtain a consensus ranking that serves as an “average” ranking, maintaining the relative importance of each brain region at the group level. 
Specifically, we propose a consensus rank-based method to test sex differences in the functional brain network topology and the topologies of individual brain regions. 
We demonstrate that females and males exhibit different centrality ranking patterns.
Further, we use the rank-biased overlap (RBO) to quantify the dissimilarity in centrality rankings between females and males, observing approximately a 10\% difference in nodal centrality consensus ranking patterns. 
To provide a more detailed analysis, we introduce a Rank Difference Index (RDI) to quantify the degree of rank difference in brain regions between sexes, identifying eight regions that exhibit the most significant rank differences between females and males.
We hypothesize that there are significant sex differences in the functional brain network topology.

\section{Material and Method}
\subsection{Dataset,  Acquisition, and Preprocessing}

The samples are collected from the Southwest University Longitudinal Imaging Multimodal (SLIM) and Gene-Brain-Behavior (GBB) projects. Detailed information about these projects can be found in prior publications \citep{liu2017longitudinal,chen2019brain}.  All participants provide written informed consent and are paid for their time and participation in the task. Approval for both projects is obtained from the Institutional Review Board of the Brain Imaging Center at Southwest University. The combined datasets consist of 1,963 right-handed, healthy young adults. Participants with excessive head movement (mean framewise displacement power $> 0.3$) during resting-state functional magnetic resonance imaging (rs-fMRI) and/or with incomplete information were excluded from the study. Finally, the dataset comprise 1,948 subjects (1,174 females, mean age: $19.76\pm0.03$; 774 males, mean age: $19.75\pm0.03$; range: $18-27$ years).

Resting-state functional magnetic resonance imaging (rs-fMRI) scans are obtained using a 3.0-T Siemens Trio MRI scanner (Siemens Medical, Erlangen, Germany). The scanning time of the rs-fMRI is 8min, using a gradient echo-planar imaging (EPI) sequence with the following parameters: slices $= 32$, repetition time (TR)/echo time (TE) $= 2000/30 \text{ms}$, flip angle $= 90^\circ$, field of view (FOV) $= 220 \times 220 \text{mm}$, and thickness/slice gap $= 3/1 \text{mm}$, and voxel size $= 3.4 \times3.4 \times 3 \text{mm}^3$. During resting-state scanning, the subjects are instructed to relax, remain awake with eyes closed, and not to think of anything in particular.

Image preprocessing is conducted using SPM12 \\ (\href{https://www.fil.ion.ucl.ac.uk/spm/}{https://www.fil.ion.ucl.ac.uk/spm/} and the Data Processing \& Analysis for Brain Imaging (DPABI) toolbox \citep{yan2016dpabi}. The preprocessing steps include discarding the first ten functional images to allow for magnetic field stabilization and participant adaptation to the scanning environment. Following this, slice timing and head motion correction are performed. The images are then spatially normalized to the Montreal Neurological Institute (MNI) template with a resampling voxel size of $= 3\times3\times3 \text{mm}^3$ and smoothed using a Gaussian kernel with a full-width at half maximum (FWHM) of $6\text{mm}$. Confounding signals, including those from white matter and cerebrospinal fluid, as well as head motion effects, are regressed out using the Friston 24-parameter model (which includes six motion parameters, their temporal derivatives, and the squares of these parameters) \citep{power2012spurious,yan2013comprehensive}. Finally, a temporal band-pass filter ($0.01 \text{Hz} - 0.1 \text{Hz}$) is applied to reduce high-frequency physiological noise.

\subsection{Construction brain network}
To construct a functional brain network for each individual, we first utilize the Power atlas \citep{power2011functional} to partition the brain into 264 brain regions of interest (ROIs), which are defined as nodes. For each ROI, time series data is acquired by averaging the signal of all voxels within that region. Following this, we compute the Pearson correlation coefficient between the time courses of each pair of ROIs to measure their connections.
To ensure the statistical significance of these connections, we apply a false discovery rate (FDR) correction with a threshold of $P < 0.001$, which aids in eliminating insignificant connections. Additionally, we discard negative connections due to their ambiguous significance in the context of functional brain networks \citep{rubinov2011weight}. These steps yield a $264 \times 264$ weighted connection matrix, \textbf{W}, for each individual. 

Moreover, the Power atlas provides a parcellation scheme for these regions, delineating 13 functional systems, including the visual network (VIN), auditory network (AUN), salience network (SN), default mode network (DMN), ventral attention network (VAN), dorsal attention network (DAN), cingulo-opercular task control (COTC), fronto-parietal task control (FPTC), sensory-somatomotor Hand (SMH), sensory-somatomotor Mouth (SMM), cerebellar (CER), and memory retrieval network (MR). These systems correspond to known large-scale brain networks that exhibit coherence during both task-related activities and rest.

Finally, label information for the brain regions is obtained from the Anatomical Automatic Labeling (AAL) atlas. For instance, PCG.R(92) refers to the AAL label ‘PCG.R,’ with ‘92’ indicating the region of interest in the Power atlas.

\subsection{Node centrality}
We consider four commonly investigated node centrality indices \citep{freeman1991centrality,van2013network}, each reflecting different aspects of network topology.

\textbf{Betweenness Centrality (BC):} BC quantifies the influence of nodes on information flow throughout the entire network by counting the times of shortest paths passing through each node.

\textbf{Degree Centrality (DC):} DC measures the local topology of a network by quantifying the total connectivity strength of nodes within a weighted network.

\textbf{Closeness Centrality (CC):} CC assesses how effectively a node can communicate and connect with other nodes in the network by summing up the reciprocals of the shortest path lengths from the node to all other nodes.

\textbf{Eigenvector Centrality (EC):} EC evaluates the importance of a node in a network by considering both its own importance and the importance of its neighboring nodes in the context of global topology.

When calculating BC and CC, the Dijkstra algorithm is used to find the shortest paths between paired nodes \citep{dijkstra1959note, ran2021novel}. While this approach may favor paths with lower correlation, which contradicts the intuitive expectation that higher correlations between nodes should facilitate communication. To address this, we apply a transformation technique to convert maximum correlation to minimum correlation, i.e., $w’_{i,j} = \frac{1}{w_{i,j}}$.

Nodal centrality scores are then rearranged in descending order and converted into integer rankings for better comprehensibility. Higher centrality scores correspond to higher ranks (i.e., smaller numerical positions). In cases where two or more nodes share identical centrality scores, they are assigned the same rank position. Consequently, The length of the nodal centrality ranking can be up to 264, denoted as  $L \leq 264$.

\subsection{Construction consensus ranking}

\begin{figure}[h]
	\centering
	\includegraphics[width=0.48\textwidth]{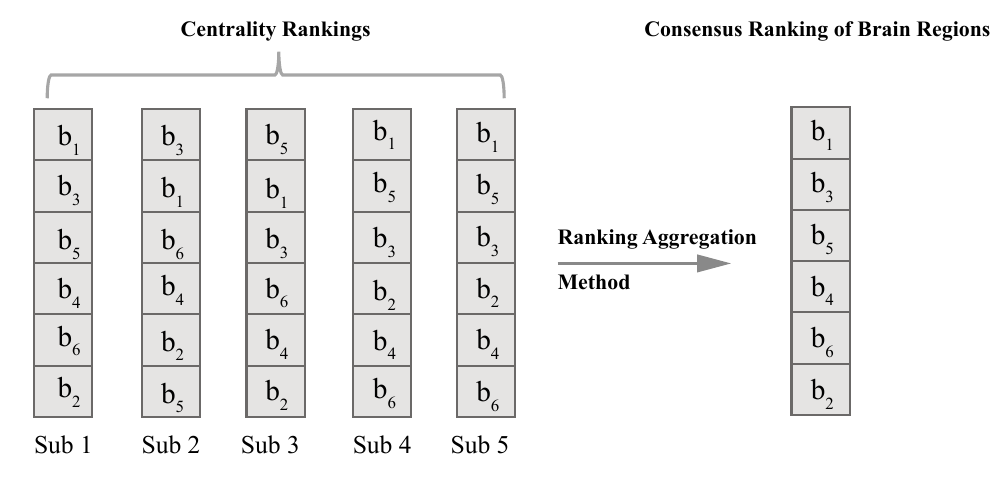}
	\caption{ Multiple centrality rankings to be aggregated. Five different subjects each rank the set of brain regions $\{b_1, b_2, b_3, b_4, b_5, b_6 \}$ based on the centrality scores}
	\label{fig 1}
\end{figure}

To accurately determine the ‘average’ ranking of brain regions and ensure that this ranking reflects the relative importance of each region, we adopt a ranking aggregation method to obtain the consensus ranking of brain regions from multiple individuals (Fig.\ref{fig 1}) \citep{dopazo2017rank,dwork2001rank}. Given the extensive length and number of ranking lists, we choose the PageRank method as our aggregation tool \citep{chen2020comparison,adali2007information}.

The implementation steps of the PageRank method are as follows. First, each individual’s ranking list is transformed into a preference matrix, $\textbf{P}$. If brain region $i$ is ranked ahead of region $j$, then $P_{i,j} = 1$; otherwise, $P_{i,j} = 0$. We then aggregate the preference matrices from all individuals to create a cumulative preference matrix, $\textbf{C}$. The elements of $C_{i,j}$ reflect the total number of brain region $i$ is preferred over brain region $j$ across all individuals.
Next, we transform the cumulative preference matrix into a directed weighted graph, which is defined as the cumulative preference graph. In this graph, each node represents a brain region, and directed edges between nodes $i$ and $j$ are weighted according to $C_{i,j}$.
To ensure that the total outgoing edge weights from each brain region sum to one, thus maintaining the stochastic nature of the PageRank algorithm, we normalize the weights in the cumulative preference graph to generate a normalized cumulative graph.
Finally, we compute the PageRank values for each region within this normalized graph. 
These PageRank values are then sorted to establish the final consensus ranking of the brain regions, reflecting their relative importance based on the aggregated preferences.

In addition, in order to verify whether the consensus ranking of brain regions effectively reflects the centrality scores of the regions, we conduct a linear regression analysis and assess the statistical significance of the results through a t-test. The analysis results suggest that the consensus ranking substantially serves as an effective indicator of region centrality scores (Fig.\ref{crs}, $R^2 = 0.96$, $P<0.001$).

\subsection{Ranking similarity measure}

To further quantify the similarity between females and males in their brain region centrality rankings, we utilize Rank-Biased Overlap (RBO) \citep{webber2010similarity}. This method employs a geometrically weighted sequence that prioritizes elements at the top of the rankings, such that higher RBO values indicate a high degree of similarity in centrality rankings across sexes, reflecting consistent functional importance within key brain regions. Conversely, lower RBO values suggest significant differences in centrality among certain critical areas.
Moreover, the adjustable parameter  $p$  within RBO allows us to tailor the influence of the top-ranked elements. A lower $p$ value highlights the most prominent brain regions, suitable when our focus is on the top-ranked areas. A higher $p$ value gives more weight to elements lower in the ranking, facilitating a more balanced evaluation of similarity. 

Numerous cognitive and behavioral studies have shown that certain key brain regions exhibit different activation and connectivity patterns between males and females \citep{tomasi2012gender}, suggesting that focusing on the ranking differences of these regions may reveal broader network topology differences \citep{van2013network}. 
Therefore, in this study, we set the RBO parameter $p$ to 0.98 to emphasize the top 20\% of brain regions (approximately the top 50 areas), identified as key areas in prior studies \citep{wang2018differentially,van2013network}. We also examine settings where $p = 0.99$ to assess the top 30\% of brain regions, which corresponds approximately to the top 80 regions.

In addition, $1 - \text{RBO}$ quantifies the dissimilarity between two rankings, where 0 represents complete similarity and values approaching 1 indicate substantial differences. This metric enables us to quantitatively analyze how sex differences manifest within brain networks.

\subsection{Consensus rank-based method}

Directly averaging multiple rankings may diminish the efficacy of statistical analyses, as this approach might obscure genuine insights into the average rankings of brain regions (Fig.\ref{item86}). Conversely, the consensus ranking more effectively captures the “average” rank of brain regions, thereby retaining more detailed ranking information. Consequently, we propose a consensus rank-based method for group comparison to more accurately delineate the importance of brain regions and their variations across sex.

This approach allows us to assess the overall differences in ranking patterns between females and males, capturing both the average discrepancy and the variability among individuals within each group. It also enables us to examine differences in ranked positions of brain regions between females and males. We will detail these aspects further below.

\textbf{Ranking patterns:} In this form, we quantify the group differences by computing the mean difference between females' consensus ranking and males' consensus ranking. Specifically, similar to the t-test, we calculate the total distance between an individual's nodal centrality ranking and the consensus ranking, which we term as the variance difference, denoted as $S^2$. The $t$ statistic is then denoted as:
\begin{equation}
\begin{gathered}
    t = \frac{D(R_{C_f}, R_{C_m})}{\sqrt{S^2(\frac{1}{N_f}+\frac{1}{N_m})}},\\
    S^2 =\frac{\sum_{i=1}^{N_f}D(R_{C_f},R_i)^2+\sum_{i=1}^{N_m}D(R_{C_m},R_i)^2}{N_f+N_m-2},
\end{gathered}  
\end{equation}
where $R_{C_f}$ and $R_{C_m}$ represent the consensus ranking for females and males respectively. $R_i$ is the nodal centrality ranking of individual $i$. $N_f$ and $N_m$ represent the sample size of females and males, respectively. $D$ represents a distance function, which is defined as $D = 1-RBO$. 

\textbf{Brain region rank:} we compute the statistic $t(v)$ for a brain region $v$ as follows:
\begin{equation}
\begin{gathered}
    t(v) = \frac{R_{C_f}(v)-R_{C_m}(v)}{\sqrt{S^2(v)(\frac{1}{N_f} + \frac{1}{N_m})}},\\
    S^2(v) = \frac{\sum_{i=1}^{N_f}(R_i(v)-R_{C_f}(v))^2+\sum_{i=1}^{N_m}(R_i(v)-R_{C_m}(v))^2}{N_f + N_m -2},
\end{gathered}
\end{equation}
where $R_{C_f}(v)$ and $R_{C_m}(v)$ represent the consensus ranking of brain region $v$ in female and male respectively. $R_i(v)$ represents the rank of brain region $v$ in the ranking of individual $i$.

Additionally, the brain region rank at the top is considered to have a greater important role than the rank at the bottom. Hence, we define a Rank Diversity Index (RDI) to measure the degree of rank differences in brain regions $v$ between females and males.
\begin{equation}
    \text{RDI}(v) = \frac{R_{C_f}(v)-R_{C_m}(v)}{\max\{R_{C_f}(v),R_{C_m}(v)\}},
\end{equation}
where $R_{C_f}(v)$ and $R_{C_m}(v)$ represent the consensus ranking positions for females and males, respectively. A positive RDI indicates that a region ranks higher in males than females, while a negative RDI suggests it ranks higher in females.

\subsection{Consensus ranking similarity based method}
To validate the effectiveness of our consensus rank-based method and gain further insights, we construct a dataset suitable for conducting a traditional two-sample t-test. The construction and analysis involve the following steps:

\textbf{Repeated grouping:} Due to unequal sample sizes between males and females, we adopt distinct sampling strategies for each sex. Males are randomly divided into two subgroups ($M_1$ and $M_2$), while females are sampled without replacement to form two subgroups ($F_1$ and $F_2$) of equal sizes as the male subgroups. This grouping procedure is repeated 1000 times to ensure robustness, resulting in $4\times1000$ datasets.

\textbf{Aggregation}: We use the PageRank method to obtain each dataset's nodal centrality consensus ranking (NCCR).

\textbf{Compute similarity:} We compute the similarity (using RBO) within each sex and between sexes, generating within-female, within-male, and between-sex similarity lists.

\textbf{Statistical analysis:} We conduct a two-sample t-test to determine if the similarity of NCCR within each sex is greater than between sexes. A significance threshold of $P < 0.05$, with a false discovery rate (FDR) controlled at $\text{FDR} < 0.05$, denotes statistical significance. Cohen's $d$ is used as an effect size index to quantify the differences between two similarity lists at the mean level. Additionally, we transform $d$ into an overlap coefficient (ovl) to assesses the entire data distribution, indicating whether the similarity between sexes outweighs the dissimilarity ($\text{ovl} > 0.5$) or vice versa.

\section{Results}
\subsection{sex differences in the nodal centrality consensus ranking patterns}

We apply a consensus rank-based method to uncover statistically significant sex differences in NCCR patterns (Table \ref{tab:addlabel}, $P < 0.05$), indicating that these differences in brain network topology are due to sex rather than individual variability. Considering the low consistency among individual centrality rankings within each sex (Fig.\ref{beforra}(a)), traditional methods that rely on mean similarity differences may not effectively distinguish between sex-based and individual variability. In contrast, the ranking aggregation method extracts subtle commonalities among individuals of the same sex, enhancing the detection of subtle sex differences in brain network topology (Fig.\ref{beforra}(b)).

\begin{table*}[H]
	\centering
	\caption{Descriptive statistics for group comparisons for BC, CC, DC, and EC.}
	\label{tab:addlabel}%
	\begin{tabular}{ccccccccc}
		\toprule
		\text{Centrality} & \text{Test Type} & \text{Sex} & \tabincell{c}{Within-sex \\(MS)} & \tabincell{c}{Between-sexes \\(MS)} & $t$ & $P$ & $d$ & \text{ovl}\\
		\midrule
		\multirow{3}[2]{*}{BC} & \text{CR} & - &- &- & 2.975 & $.001$&-&-\\
		\cmidrule{2-9} &  \multirow{2}[2]{*}{CRS} &\text{Female} & 0.90 & 0.86 & 87.69 &$\sim 0 $  & 3.92  & 0.0509  \\
		&                                     & \text{Male} & 0.92 & 0.86 & 129.19 & $\sim 0 $ & 5.77  & 0.0062 \\
		\midrule
		\multirow{3}[2]{*}{CC} &  \text{CR} & - &- &- & 2.79 & $.002$&-&-\\
		\cmidrule{2-9} &  \multirow{2}[2]{*}{CRS} &\text{Female} & 0.91 & 0.86 & 91.32 & $\sim 0 $   & 3.98  & 0.0429  \\
		&& \text{Male} & 0.91 & 0.86 & 92.06 & $\sim 0 $   & 4.57  & 0.0341  \\
		\midrule
		\multirow{3}[2]{*}{DC} &  \text{CR} & - &- &- & 2.599 & $.004$&-&-\\
		\cmidrule{2-9} &  \multirow{2}[2]{*}{CRS} &\text{Female} & 0.93 & 0.90 & 88.93 & $\sim 0 $ & 4.08  & 0.0439 \\
		&& \text{Male} & 0.94 & 0.90 & 102.34 & $\sim 0 $  & 4.12  & 0.0428 \\
		\midrule
		\multirow{3}[2]{*}{EC} &  \text{CR} & - &- &- & 2.77 & $.002$&-&-\\
		\cmidrule{2-9} &  \multirow{2}[2]{*}{CRS} & \text{Female} & 0.91 & 0.88 & 30.94 & $4.23\times 10^{-172}$ & 1.38  & 0.5196 \\
		&& \text{Male} & 0.91 & 0.88 & 45.51 & $1.02\times 10^{-273}$ & 1.9   & 0.3297 \\
		\bottomrule
	\end{tabular}
	
	{\raggedright \small Note: CR = Censensus Rank-based Method, CRS = Censensus Ranking Similarity-based Method, MS = the average RBO, $d$ = the effect size of Cohen's $d$ (mean differences), $\text{ovl} =$ overlap coefficient.\par}
\end{table*}%
\FloatBarrier

\begin{figure*}[h]
	\centering
	\includegraphics[width=0.8\textwidth]{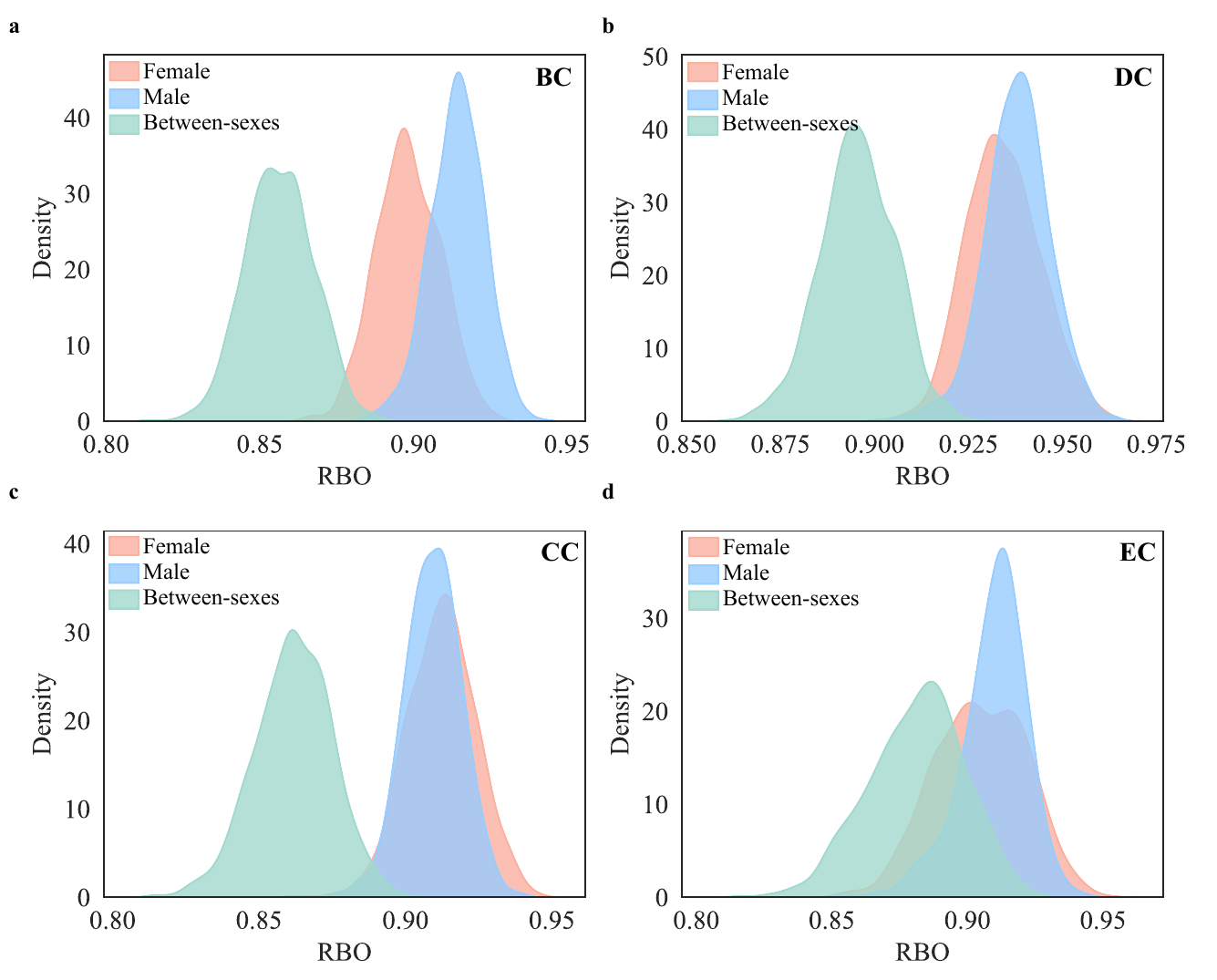}
	\caption{Kernel density plots of similarity in within-females, within-males, and between-sexes on BC, DC, CC, and EC respectively. Shown the results of RBO in $p = 0.98$.}
	\label{fig 2}
\end{figure*}

Moreover, we also conduct the consensus rank similarity-based method to reveal that the within-sex similarity of consensus rankings is significantly higher than the between-sexes similarity (Table.\ref{tab:addlabel} and Fig.\ref{fig 2}, $P < 0.05$ and $ \text{FDR} < 0.05$). Negative effect sizes ($d$) imply higher similarity for between-sexes, while positive $d$ values indicate higher similarity for within-sex. Effect sizes range from 1.38 to 5.77, suggesting higher similarity within-sex for all centrality indicators. Furthermore, employing the overlap coefficient (ovl) also reveals significant sex differences in NCCR, especially in BC, CC, and DC. 
In addition, we conduct the same analysis process using RBO with $p = 0.99$, observing consistent results (Table \ref{tab:099} and Fig.\ref{fig s1}).
These findings suggest that within the brain network’s topology, males and females each have distinct ranking patterns, with individuals of the same sex showing greater consistency in the importance of nodes within the brain functional network, while this consistency is lower between the sexes.

Finally, we observe that the degree of sex differences in the patterns of NCCR is approximately 10\% ($\text{dissimilarity} = 1 - \text{RBO}$) for DC, EC, BC, and CC. The smallest differences are observed in DC and EC, while the largest differences are found in BC and CC. This suggests that differences in global network topology, as reflected by EC, BC, and CC, are greater than those in local network topology (DC). Given the RBO parameter set at 0.98, these significant sex differences may be attributed to variations in the consensus ranking of the top 50 brain regions. Our next step is to examine sex differences in these specific brain regions.

\subsection{Sex differences in the Top 50 brain regions}

We use the consensus rank-based method to examine the rank differences in brain regions within the top 50 between females and males. At the same time, we demonstrate that this method is more sensitive and robust in detecting differential brain regions, particularly in small sample sizes (Fig. \ref{overlapresults}). 

\begin{figure*}[H]
	\centering
	\includegraphics[width=0.8\textwidth]{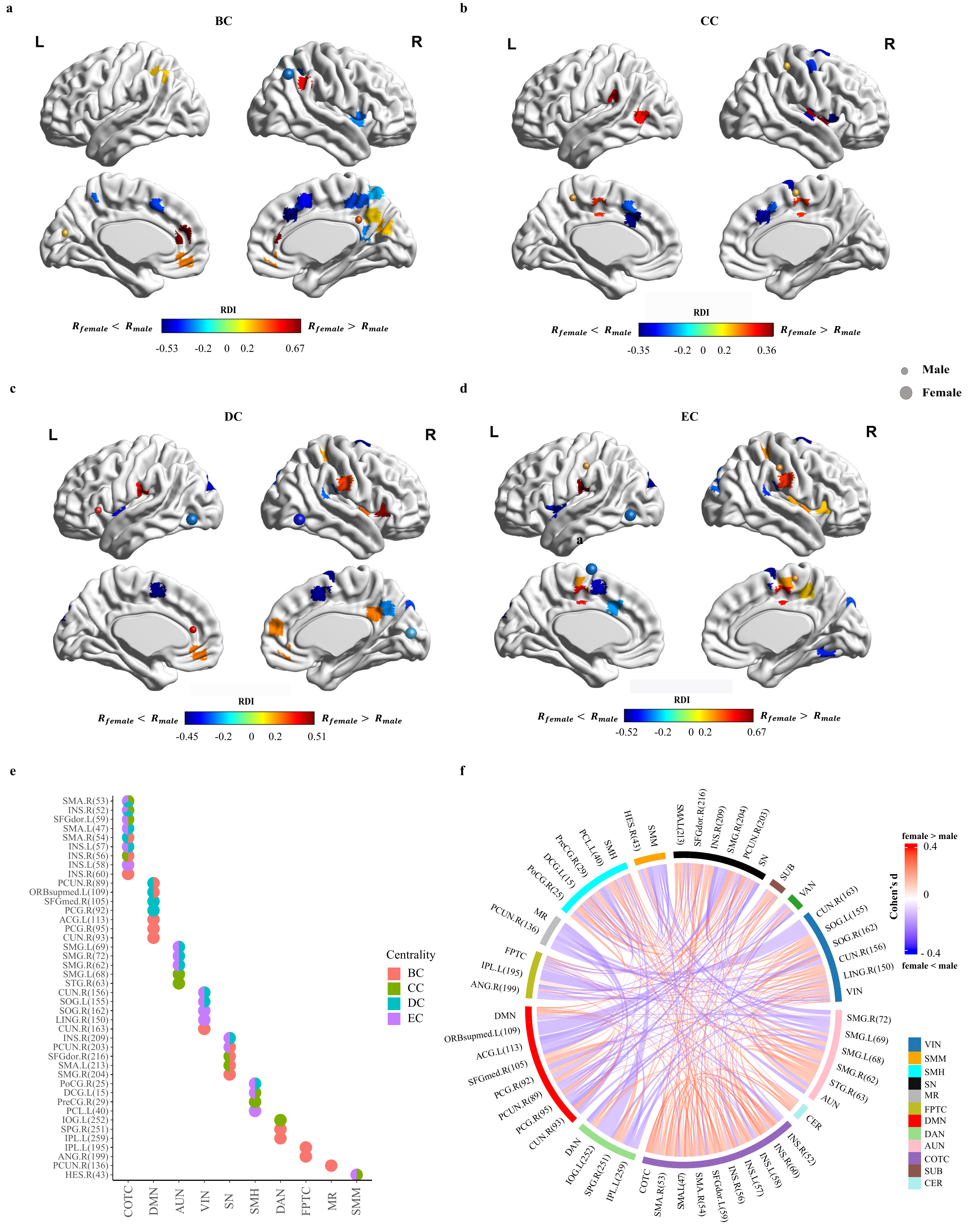}
	\caption{Results for differences in brain regions. \textbf{(a)-(d)} Spatial maps for each of 4 centralities including BC, CC, DC, and EC, a consensus rank-based method ($P \leq 0.05$) and rank difference index (RDI) are used to seek rank differences between females and males ( red = higher rank in females; blue = higher rank in males; darker = larger rank differences). Only rank differences (RDI) larger than $\pm 0.2$ are shown. Small circles represent brain regions only rank in the Top 50 for males with $ \text{RDI} \geq 0.2$, while large circles represent brain regions only rank in the Top 50 for females with $ \text{RDI} \leq -0.2$. The color indicates the value of RDI. \textbf{(e)} Compositions in the indicator set based on brain regions with significant sex differences. These are sorted by the number of brain regions within each subsystem that show significant gender differences (i.e., the brain regions in (a)-(d)). \textbf{(f)} Chord diagram for individual connections. Color and line thickness represent the effect sizes of sex on the strength of connections (red = stronger in female; blue = stronger in males; thicker = larger effect size). Only effect sizes (Cohen's $d$) larger than $\pm 0.2$ are shown. Nodes refer to the brain regions shown in (a)-(d).}
	\label{fig 3}
\end{figure*}

Fig. \ref{fig 3}(a)-(d) and Table \ref{tab:S1} display the brain regions meeting the criteria of $P \leq 0.05$ and $| \text{RDI}|\geq 0.2$. We first find that forty-two brain regions exhibit significant sex differences and rank differences. 
Among them, a large body of brain regions are identified in EC and BC, followed by DC and CC, suggesting regions’ BC and EC are more sensitive to sex differences.
Meanwhile, most brain regions show consistent directions of sex differences across different centrality measures, regardless of whether BC, DC, or EC is used; that is, one sex typically ranks higher than the other across all measures.
On the other hand, the right Precuneus (PCUN.R(89)) and PCUN.R(203) are exceptions. 
PCUN.R(89) ranks higher in females for DC, but higher in males for BC, with similar ranking differences for both measures. 
PCUN.R(203) ranks higher in females for BC, but higher in males for EC.

Further narrowing down to the eight brain regions with the largest sex differences ($| \text{RDI}|\geq 0.5$), we find the involvement of three centrality indicators: BC, EC, and DC. In BC,females display higher ranks in the right dorsolateral superior frontal gyrus (SFGdor.R(216)) and the right supplementary motor area (SMA.R(54)) compared to males.
Conversely, males show higher ranks in the left anterior cingulate and paracingulate gyri (ACG.L(113)) and the right supramarginal gyrus (SMG.R(204)). In EC, females show higher ranks in the SMA.R(53), left insula (INS.L(57)), and left superior occipital gyrus (SOG.L(155)), while males exhibit a higher rank in the SMG.L(69). Regarding DC, females rank lower than males in the INS.R(209).

\FloatBarrier
\noindent

\subsection{Detailed analyses on the brain regions with significant rank differences between sexes}

In order to gain detailed insights into the brain regions with significant sex differences, we first investigate the functional system ascription of these regions and compare the composition across four centrality measures (Fig. \ref{fig 3}e), given that different centrality indicators reflect different dimensions of brain region characteristics (Fig. \ref{figS2}). Females show higher rankings in the COTC and VIN, while males rank higher in the DMN and AUN. Specifically, the brain regions with significant differences are mainly associated with subsystems such as COTC, DMN, AUN, VIN, and SN.
Among these, brain regions in the COTC and SN exhibit significant sex differences across all four centrality measures, particularly in the SMA and INS areas of the COTC. 
The brain regions in AUN are primarily concentrated in the SMG and show more significant differences in the CC, DC, and EC dimensions. 
Brain regions in the VIN display more differences in the BC, DC, and EC dimensions. 
The DMN regions show significant differences only in the BC and DC dimensions.
These results suggest that some subsystems, such as COTC, AUN, and VIN, reflect functional differences between male and female brains across multiple dimensions, while others, like DMN and DAN, involve fewer dimensions and may reveal sex differences in only a single centrality measure, as seen in FPTC and MR.

In addition, to better understand the underlying mechanisms of the observed sex differences in brain region ranks, we investigate the differences in functional connectivity (FC) in these regions by conducting an analysis of covariance (ANCOVA), using age as a covariate. 
As illustrated in Fig. \ref{fig 3}(f), substantial significant sex differences in FC strength are observed between these brain regions ($P<0.001$, $\text{FDR} <0.05$, and Cohen’s $|d|\geq 0.2$). Females exhibit stronger connections within systems, particularly within the COTC, whereas males display stronger connections between systems, especially between the DMN, FPTC, MR, and other subsystems.

In conjunction with Fig. \ref{fig 3} (a)-(e), females show higher rankings in COTC-related brain regions in DC and EC, likely due to their strong intra-system and inter-system connectivity (e.g., COTC-SN, COTC-VIN, and COTC-SMH). Additionally, these systems display stronger interactions with other connected systems, such as VIN-AUN. These findings suggest that the COTC system is more active in the female brain, which may influence the global topology of the brain network. In contrast, males show higher rankings in some DMN regions in BC and DC, likely due to the greater number of cross-system connections in these regions.

\section{Discussion \& Limitation}
In this study, we investigate sex-related effects in human functional brain network using a consensus rank-based approach. Our results reveal statistically significant differences in NCCRs between females and males, with these differences primarily attributable to biological sex rather than individual variability. Specifically, eight key brain regions exhibited the most pronounced rank differences between the sexes. Related to the functional system and functional connectivity, females have higher rankings in regions belonging to COTC and VIN, which show stronger within-system connectivity. Whereas males have higher rankings in regions associated with the DMN and AUN that display stronger between-system connectivity. 
These findings underscore the importance of considering biological sex in brain network studies and provide a foundation for further exploration of sex-specific network characteristics.

Our findings support the hypothesis that there are statistically significant sex differences in functional brain network topology, as revealed through the analysis of four nodal centrality indicators. While previous studies have primarily examined sex differences at the level of mean nodal centrality values \citep{zuo2012network,tian2011hemisphere}, our research extends this by focusing on consensus rankings. Shifting from mean values to consensus rankings offers several advantages, including the ability to capture common characteristics among individuals of the same sex, reflect population-level consensus in brain region rankings, and facilitate cross-sex comparability. Notably, our results demonstrate that the observed differences in functional brain network topology are attributable to biological sex rather than individual variability, challenging prior assertions that these differences are rooted in individual factors \citep{cahill2006sex,kaczkurkin2019sex,cosgrove2007evolving,grabowska2017sex}.

Notably, this study is the first to investigate sex differences in brain network topology using a consensus rank-based method, revealing that the degree of difference in functional brain networks is approximately 10\% based on nodal centrality consensus rankings. This finding aligns with previous reports suggesting that the male brain is roughly 11\% larger than the female brain, yet differences in anatomical structure and lateralization are only about 1\% after controlling for brain size \citep{eliot2021dump}. These results imply that the consensus ranking method can detect subtle functional connectivity differences that are not solely dependent on overall brain size. Future research should further explore the influence of brain size on these findings.

At the brain region level, our results highlight the superiority of the consensus rank-based method proposed in this study, compared to traditional approaches such as the mean difference-based method and the mean rank-based method (Fig. \ref{overlapresults}). By integrating this new method with the rank difference index, we identified 42 brain regions with significant sex differences across four centrality metrics. These regions are consistent with those identified in previous studies using seed-based, functional connectivity-based, and graph theory-based methods that have also demonstrated significant sex differences \citep{allen2011baseline,biswal2010toward,zhang2018functional,de2019multilevel}. For example, seed-based methods using the PCG in the DMN as a seed node found stronger connections in females \citep{biswal2010toward}, aligning with our finding of higher BC in females at the PCG. This suggests that while the new method is comparable to previous methods, it offers higher sensitivity in detecting sex differences. 
It should be noted that different methods may capture distinct functional characteristics within the same brain regions, reflecting diverse aspects of connectivity and function between females and males. 

Furthermore, by narrowing the rank difference range, we identify eight hub brain regions with the most significant sex differences in population consensus rankings. These key regions closely resemble the hub regions identified in previous resting-state \citep{van2013network,bell2016subcortical,hagmann2008mapping,van2011rich,zuo2012network} and task-related \citep{dosenbach2008dual,xu2020sex,dosenbach2007distinct} brain network studies. These regions are crucial for the processing and transmission of information within the brain, as highlighted in \citep{van2013network}. Our findings further support the hypothesis that females and males might employ different cognitive mechanisms to accomplish the same tasks or achieve similar behavioral outcomes \citep{de2004minireview, cahill2006sex, grabowska2017sex, gur2017complementarity, mccarthy2016multifaceted,xu2020sex}. However, questions remain about the specific strategies employed by females and males to regulate their behavioral performance and how brain neuronal activity is linked to these behaviors. These critical areas warrant further in-depth exploration to uncover the underlying mechanisms of sex differences in cognitive processing.

Additionally, our results highlight the importance of COTC and VIN in females, which may reflect their advantages in task control, attention maintenance, and visual perception \citep{dosenbach2007distinct,sheffield2015fronto,wallis2015frontoparietal}. The significant differences observed in COTC across the four centrality metrics underscore its crucial role in integrating information and coordinating task execution \citep{rawls2022resting}. Conversely, the higher rankings of brain regions in the DMN and AUN in males may be indicative of their advantages in self-related thoughts \citep{bertossi2017transcranial} and auditory processing \citep{krizman2020sex}. The differences in DMN and AUN regions are primarily reflected in the BC and DC dimensions, aligning with the characteristics of information processing and transmission in these systems in males.
Further analysis shows that females demonstrate stronger within-system connectivity, while males exhibit stronger inter-system connectivity, consistent with previous findings \citep{filippi2013organization, smith2014characterizing,biswal2010toward,ritchie2018sex}. Notably, we find that females have stronger connectivity within the COTC, which was not clearly identified in previous studies focusing solely on functional connectivity. We speculate that the higher BC, CC, DC, and EC rankings of COTC regions in females may be attributed to their robust within-system and between-system connections, while the higher BC and DC rankings of DMN regions in males may be related to their extensive between-system connections.
These results suggest that our proposed method, by incorporating multidimensional centrality analysis, provides more detailed detection of functional system differences than traditional methods, demonstrating enhanced sensitivity and specificity across multiple centrality dimensions.

Indeed, our study provides valuable insights into sex differences in functional brain network topology and underscores the effectiveness of the consensus rank-based method for group comparisons. However, several limitations should be considered when interpreting our findings. First, we use the Power Atlas and Pearson correlation to construct brain networks, but recent research show that thresholding and weighting schemes \citep{buchanan2020effect, de2013estimating, roberts2017consistency} and the number of nodes and edges \citep{mheich2020brain} significantly impact network topology. Future studies should explore how varying the number of nodes, thresholds, and weighting schemes affects sex differences in brain network topology. Second, our method treats each brain region’s of consensus ranking as unique, without accounting for the possibility of ties or equal ranks. Investigating tied rankings could offer a more nuanced understanding of brain region importance. Finally, while our ranking aggregation method effectively compares males and females at the population level and reduces individual variability, other factors—such as brain size, cultural backgrounds, and socioeconomic factors—may also influence sex differences in brain structure and function \citep{eliot2021dump,kiesow202010}. Future studies should incorporate these factors to better understand the role of ranking aggregation method in uncovering population differences.

\section{Conclusion}
In summary, we propose a novel consensus rank-based approach to assess sex differences in functional brain network topology, focusing on nodal centrality rankings. Our method reveals statistically significant differences in nodal centrality consensus rankings between females and males, attributing these differences to biological sex rather than individual variations. The study highlights the importance of biological sex in brain network organization, with our method pinpointing eight key brain regions with the most pronounced rank differences.
While our findings provide robust evidence of sex differences, they are based on a specific dataset, which may limit their generalizability. Future research should validate this approach across diverse populations and integrate it with other network analysis techniques to enhance our understanding of brain network organization. Despite these limitations, the consensus rank-based method shows promise for exploring group differences beyond sex, potentially enriching insights into network neuroscience.

\section{Acknowledgements}

\section{Author contributions}
Wenyu Chen: Formal analysis, Methodology, Visualization, Writing–original draft, Writing–review \& editing. Ling Zhan: Formal analysis. Yunsong Luo: Formal analysis. Jiang Qiu: Resources. Tao Jia: Conceptualization, Supervision, Writing – original draft, Writing–review \& editing.

\section{competing financial interests:}
The authors declare no competing financial interests.

\section{Bibliography}

\bibliographystyle{cas-model2-names}
\bibliography{referen}

\appendix
\section{My Appendix}
\renewcommand{\thefigure}{S\arabic{figure}}
\setcounter{figure}{0}

\begin{figure*}[htbp]
	\centering
	\includegraphics[width=1\textwidth]{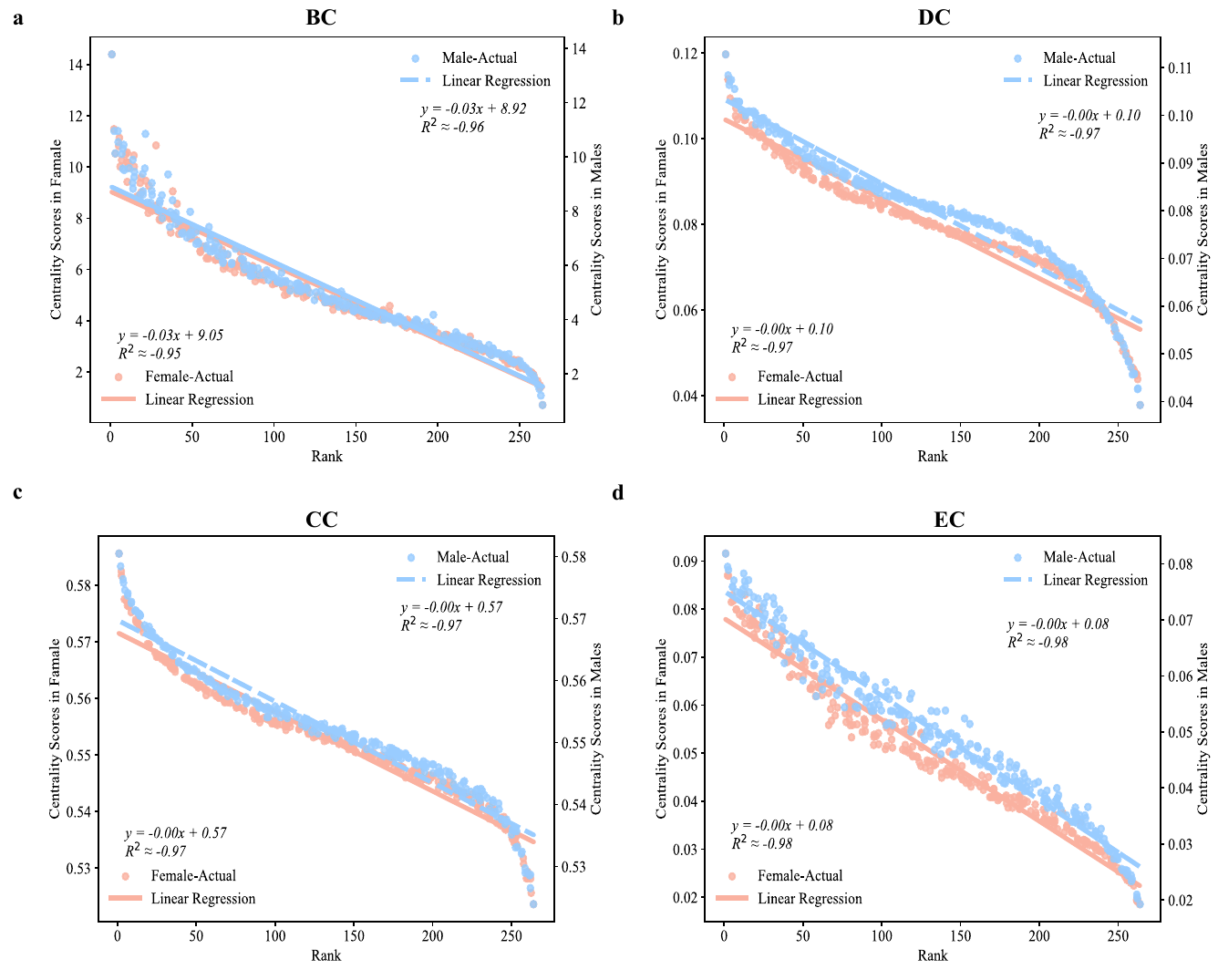}
	\caption{Regression for nodal centrality consensus ranking and nodal centrality scores on BC, CC, DC, and EC respectively. Regression analysis is performed with consensus rankings as the independent variable ($x$) and centrality scores as the dependent variable ($y$). The resulting regression lines and their corresponding equations are shown in the figure, with all $R^2$ values greater than 0.95, indicating a strong correlation between the rankings and the centrality scores.}
	\label{crs}
\end{figure*}

\clearpage
\begin{figure*}[htbp]
	\centering
	\includegraphics[width=1\textwidth]{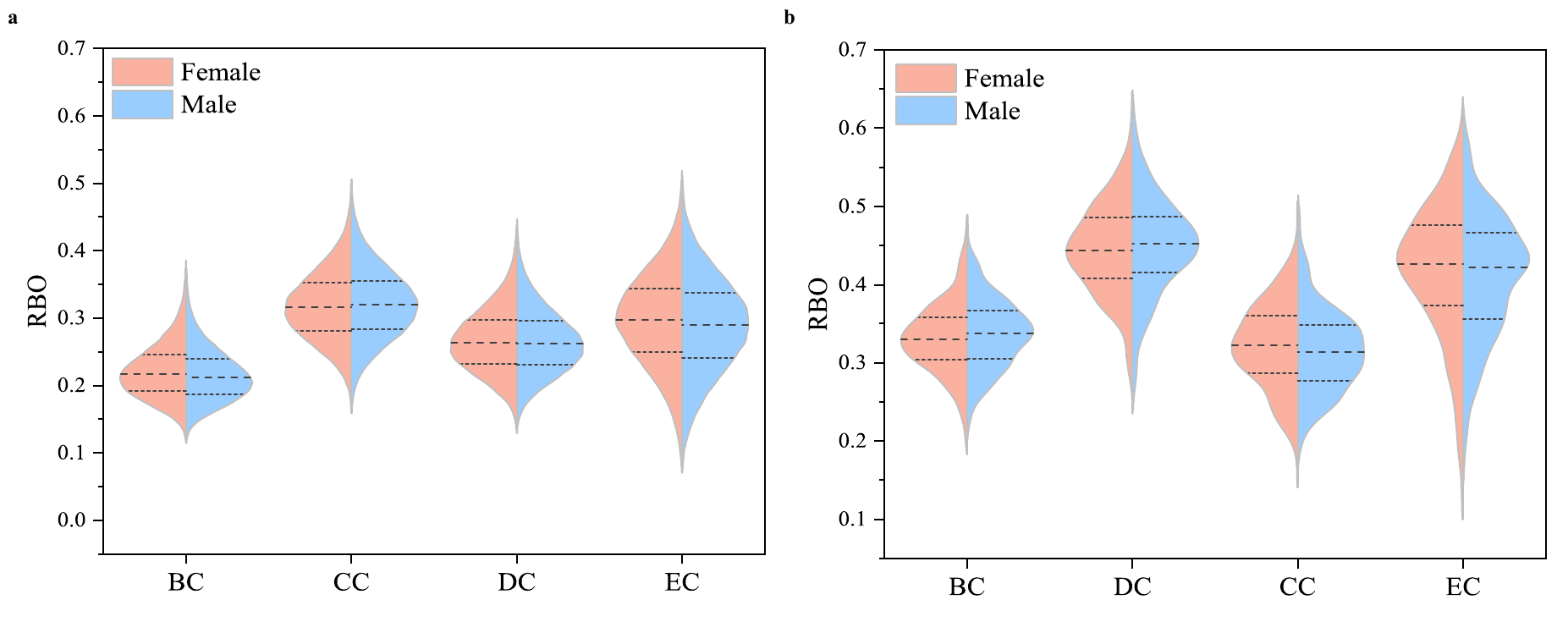}
	\caption{Similarity among individuals for each sex. \textbf{(a)} The similarity among individual rankings for each sex, showing low consistency among individual centrality rankings within each sex. \textbf{(b)} The similarity between individual ranking and consensus ranking, demonstrating that the consensus ranking reflects a moderate level of consistency with individual rankings.}
	\label{beforra}
\end{figure*}

\clearpage
\begin{figure*} 
	\centering
	\includegraphics[width=1\textwidth]{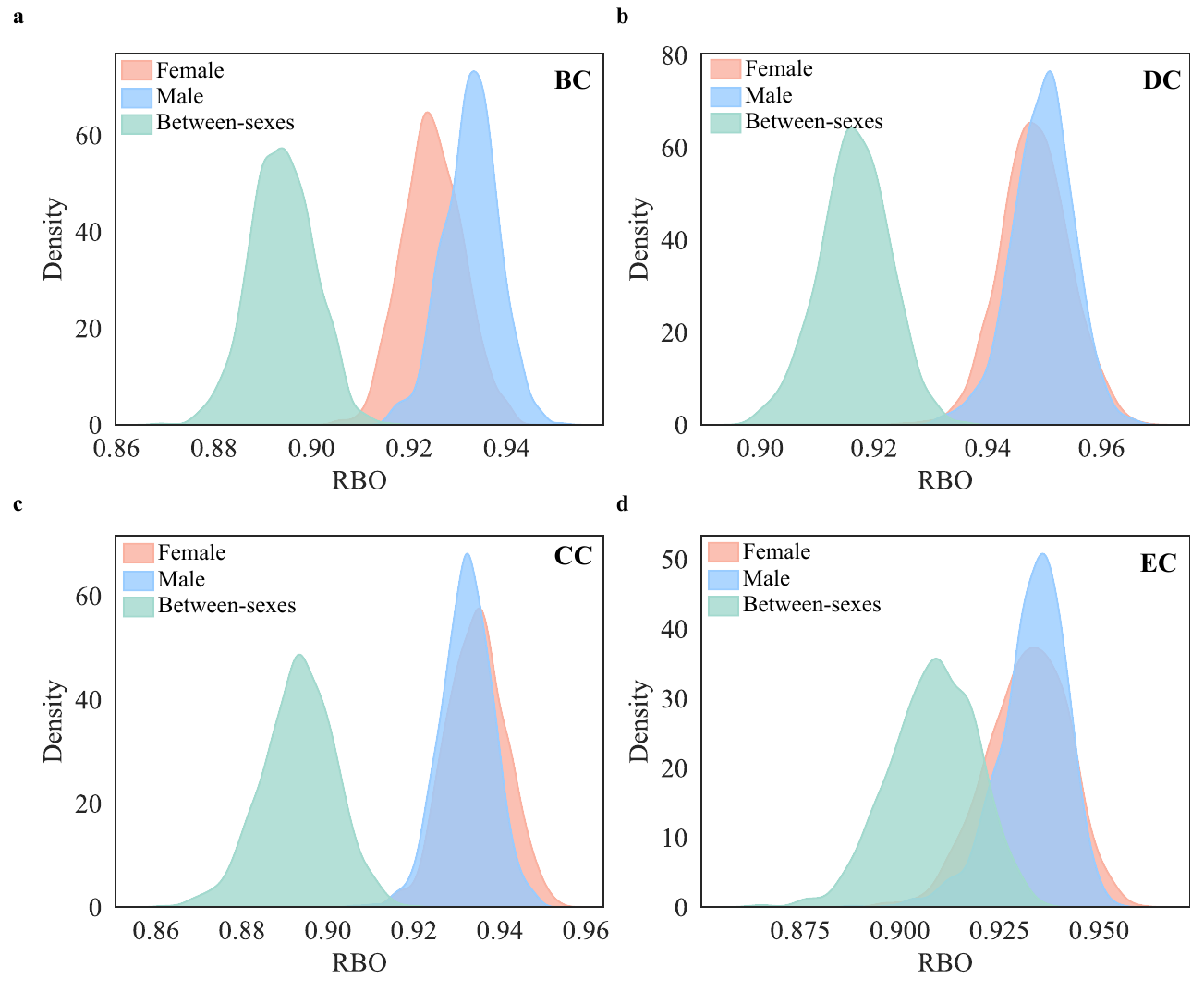}
	\caption{Kernel density plots of similarity in within-females, within-males, and between-sexes. Shown the results of RBO in $p = 0.99$. d represents the effect size of Cohen's d (mean differences). ovl = Overlap coefficient. All mean differences were statistically significant at $P < 5.10\times 10^{-149}$. }
	\label{fig s1}
\end{figure*}

\clearpage
\begin{figure*}[htbp]
	\centering
	\includegraphics[width=0.8\textwidth]{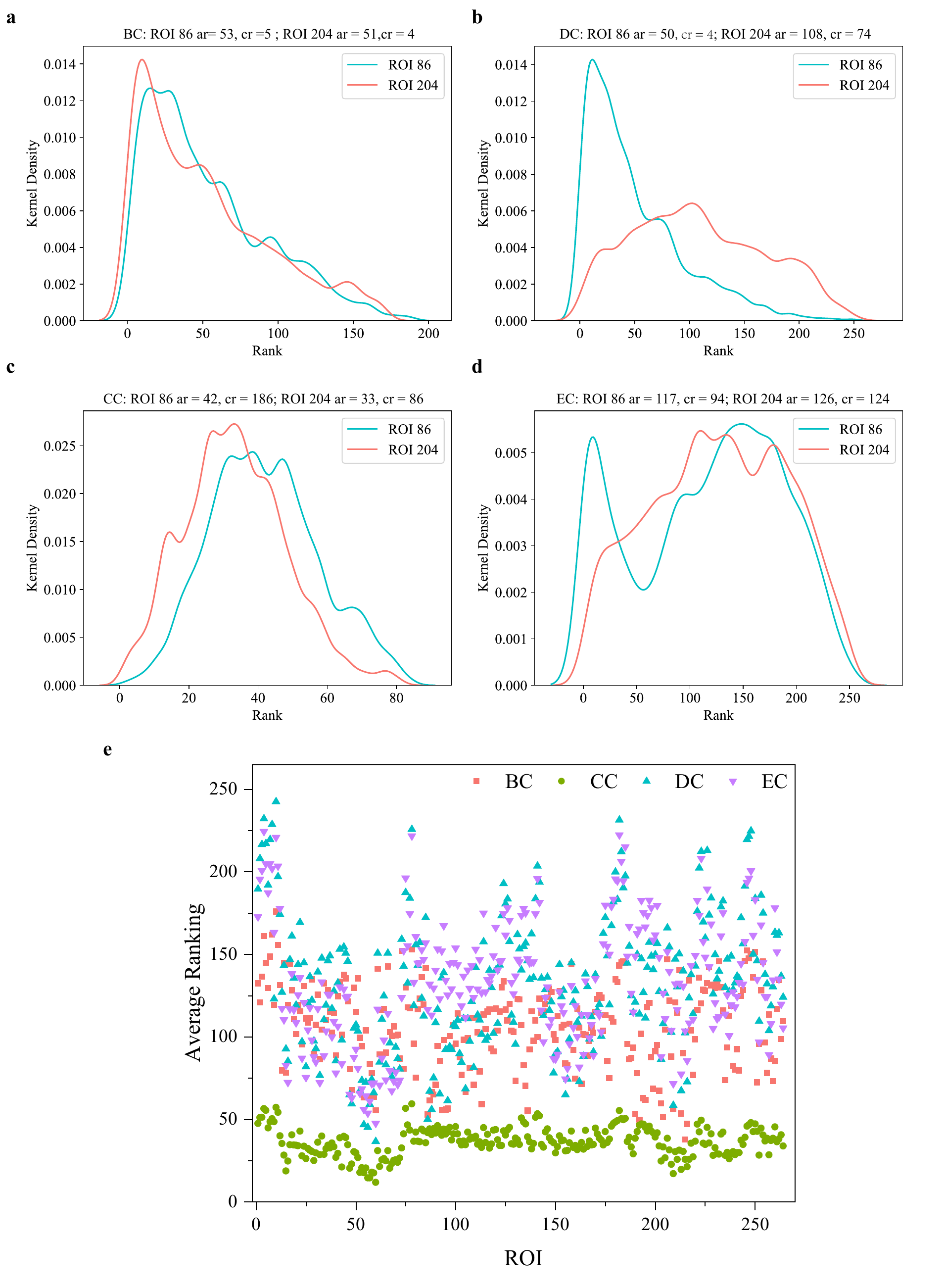}
	\caption{Compare average ranking (ar) with consensus ranking (cr) for the 86 and 204 brain regions in BC, CC, DC, and EC respectively. \textbf{(a-b)} The average ranking may not reflect the overall level. \textbf{(c)} The average ranking may not consider the relative importance. \textbf{(d)} The average ranking similar consensus ranking.  \textbf{(e)} The average ranking lose the ranking information.}
	\label{item86}
\end{figure*}

\clearpage

\begin{figure*}[htbp]
	\centering
	\includegraphics[width=1\textwidth]{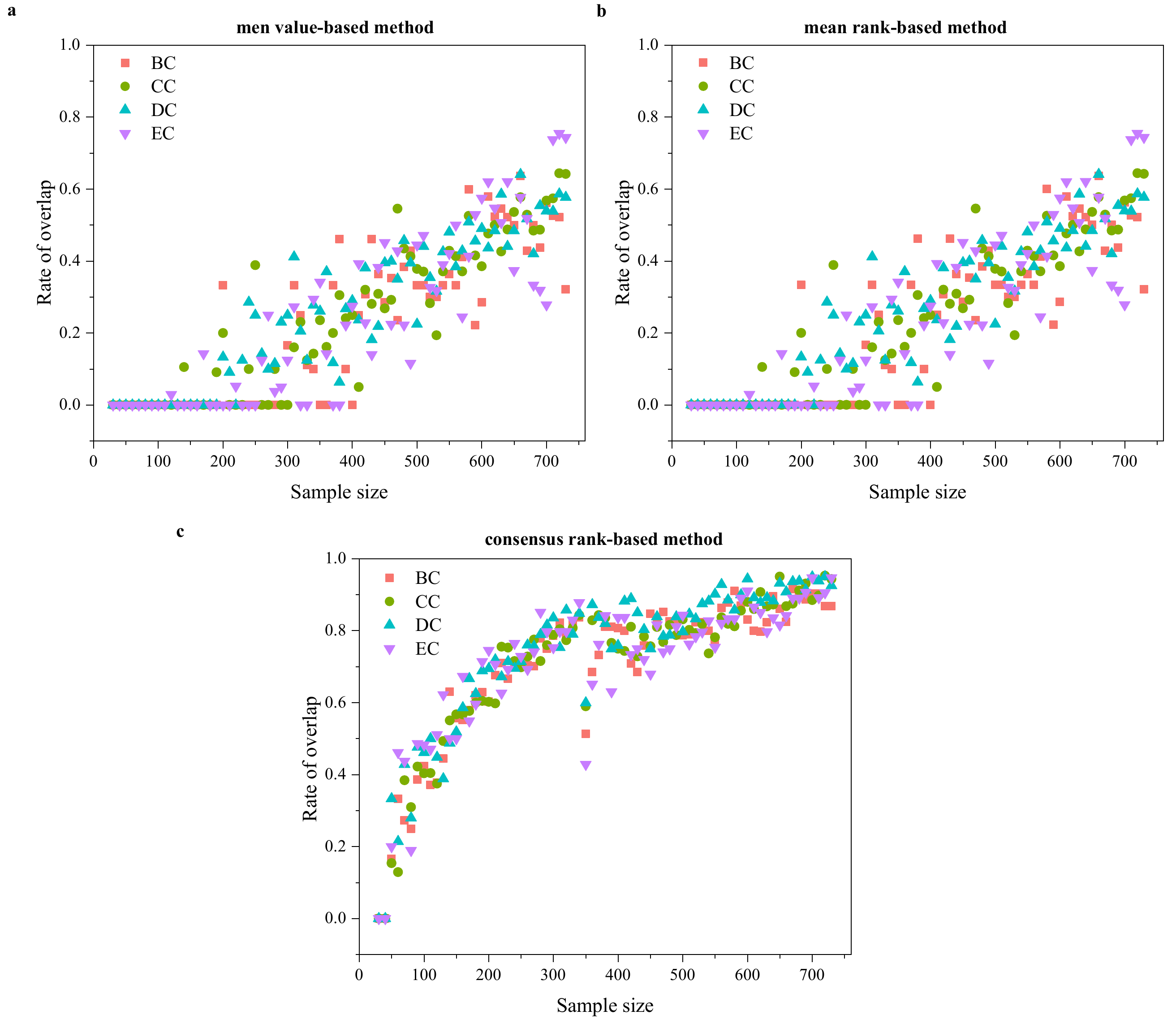}
	\caption{The variation of the rate of overlap of brain regions with significant sex differences as the sample size changes. \textbf{(a)} mean value-based method; \textbf{(b)} mean rank-based method; \textbf{(c)} consensus rank-based method. }
	\label{overlapresults}
\end{figure*}

\clearpage
\begin{figure*}
	\centering
	\includegraphics[width=1\textwidth]{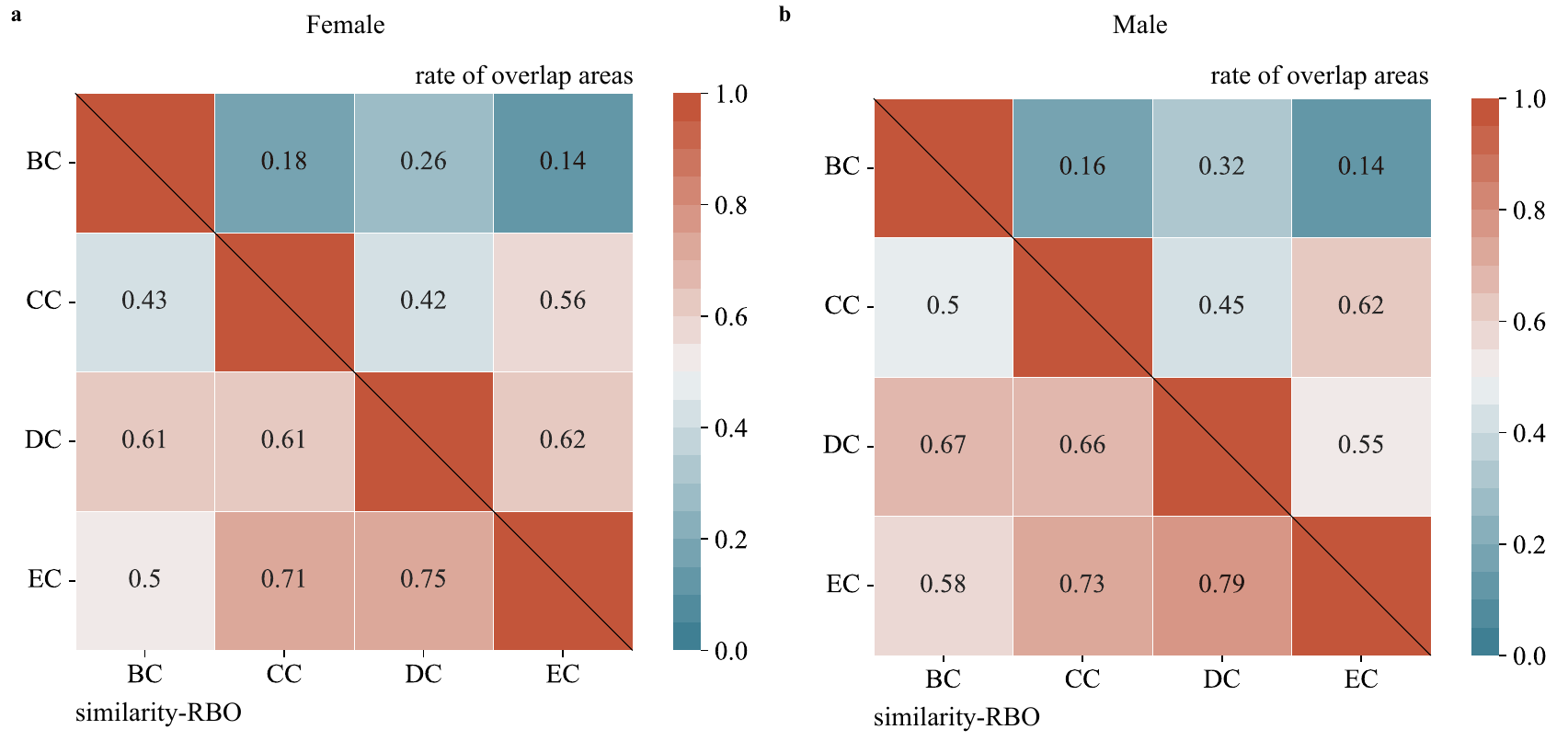}
	\caption{The similarity and the rate of overlap areas among BC, CC, DC, and EC in the Top 50 consensus rankings for each sex. Upper triangle represents the rate of overlap areas, namely the proportion of the number of identical areas between any indicators and the overall areas. Lower triangular represents the similarity among the four indicators based on RBO with $p = 0.98$.}
	\label{figS2}
\end{figure*}

\clearpage
\setcounter{table}{0}
\renewcommand{\thetable}{S\arabic{table}}
\begin{table*}[htbp]
  \centering
  \caption{Descriptive statistics for group comparisons for BC, CC, DC, and EC.}
  \label{tab:099}%
  	\begin{tabular}{ccccccccc}
  	\toprule
  	\text{Centrality} & \text{Test Type} & \text{Sex} & \tabincell{c}{Within-sex \\(MS)} & \tabincell{c}{Between-sexes \\(MS)} & $t$ & $P$ & $d$ & \text{ovl}\\
  	\midrule
  	\multirow{3}[2]{*}{BC} & \text{CR} & - &- &- & 2.66 & $.004$&-&-\\
  	\cmidrule{2-9} &  \multirow{2}[2]{*}{CRS} &\text{Female} & 0.925 & 0.89 & 109.2 &$\sim 0 $  & 4.88  & 0.0149  \\
  	&                                     & \text{Male} & 0.935 & 0.89 & 142.54 & $\sim 0 $ & 6.37  & 0.0016 \\
  	\midrule
  	\multirow{3}[2]{*}{CC} &  \text{CR} & - &- &- & 2.34 & $.009$&-&-\\
  	\cmidrule{2-9} &  \multirow{2}[2]{*}{CRS} &\text{Female} & 0.93 & 0.89 & 120.2 & $\sim 0 $   & 5.37  & 0.0810  \\
  	&& \text{Male} & 0.932 & 0.89 & 119.27 & $\sim 0 $   & 5.33  & 0.0094  \\
  	\midrule
  	\multirow{3}[2]{*}{DC} &  \text{CR} & - &- &- & 2.63 & $.004$&-&-\\
  	\cmidrule{2-9} &  \multirow{2}[2]{*}{CRS} &\text{Female} & 0.95 & 0.92 & 116.97 & $\sim 0 $ & 5.23  & 0.0110 \\
  	&& \text{Male} & 0.95 & 0.92 & 127.81 & $\sim 0 $  & 5.71  & 0.0084 \\
  	\midrule
  	\multirow{3}[2]{*}{EC} &  \text{CR} & - &- &- & 2.71 & $.003$&-&-\\
  	\cmidrule{2-9} &  \multirow{2}[2]{*}{CRS} & \text{Female} & 0.93 & 0.91 & 50.2 & $\sim 0 $ & 2.24  & 0.2765 \\
  	&& \text{Male} & 0.93 & 0.91 & 57.08 & $\sim 0$ & 2.55  & 0.206 \\
  	\bottomrule
  \end{tabular}
  
 {\raggedright \small Note: CR = Censensus Rank-based Method, CRS = Censensus Ranking Similarity-based Method, MS = the average RBO, $d$ = the effect size of Cohen's $d$ (mean differences), $\text{ovl} =$ overlap coefficient.\par}
 \end{table*}%

\clearpage
\begin{table}[htbp]
  \centering
  \caption{Ranking differences in the Top brain regions.}
    \begin{tabular}{llrrrrrrrrrrrr}
    \toprule
    \multicolumn{1}{c}{\multirow{2}[4]{*}{\textbf{Label (AAL)}}} & \multicolumn{1}{c}{\multirow{2}[4]{*}{\textbf{System}}} & \multicolumn{3}{c}{\textbf{BC}} & \multicolumn{3}{c}{\textbf{CC}} & \multicolumn{3}{c}{\textbf{DC}} & \multicolumn{3}{c}{\textbf{EC}} \\
\cmidrule(r){3-5}\cmidrule(r){6-8}\cmidrule(r){9-11}\cmidrule(r){12-14}         &       & \multicolumn{1}{c}{\textbf{$R_f$}} & \multicolumn{1}{c}{\textbf{$R_m$}} & \multicolumn{1}{c}{\textbf{RDI}} & \multicolumn{1}{c}{\textbf{$R_f$}} & \multicolumn{1}{c}{\textbf{$R_m$}} & \multicolumn{1}{c}{\textbf{RDI}} & \multicolumn{1}{c}{\textbf{$R_f$}} & \multicolumn{1}{c}{\textbf{$R_m$}} & \multicolumn{1}{c}{\textbf{RDI}} & \multicolumn{1}{c}{\textbf{$R_f$}} & \multicolumn{1}{c}{\textbf{$R_m$}} & \multicolumn{1}{c}{\textbf{RDI}} \\
    \midrule
    DCG.L(15) & SMH   & -     & -     & -     & 9     & 7     & 0.240  & -     & -     & -     & 32    & 16    & 0.49  \\
    PoCG.R(25) & SMH   & -     & -     & -     & -     & -     & -     & 32    & 25    & 0.21  & 18    & 12    & 0.32  \\
    PreCG.R(29) & SMH   & -     & -     & -     & 24    & 31    & -0.23  & -     & -     & -     & -     & -     & - \\
    PCL.L(40) & SMH   & -     & -     & -     & -     & -     & -     & -     & -     & -     & 22    & 15    & 0.30  \\
    HES.R(43) & SMM   & -     & -     & -     & 14    & 9     & 0.36  & -     & -     & -     & 35    & 25    & 0.29  \\
    SMA.L(47) & COTC  & -     & -     & -     & -     & -     & -     & 11    & 19    & -0.43  & 6     & 10    & -0.44  \\
    INS.R(52) & COTC  & -     & -     & -     & 5     & 3     & 0.36  & 14    & 10    & 0.29  & 8     & 5     & 0.32  \\
    SMA.R(53) & COTC  & -     & -     & -     & 14    & 20    & -0.32  & 20    & 36    & -0.45  & 10    & 20    & -0.51  \\
    SMA.R(54) & COTC  & 19    & 32    & -0.40  & -     & -     & -     & 3     & 5     & -0.43  & -     & -     & - \\
    INS.R(56) & COTC  & 18    & 25    & -0.28  & 3     & 4     & -0.35  & -     & -     & -     & -     & -     & - \\
    INS.L(57) & COTC  & -     & -     & -     & -     & -     & -     & 6     & 10    & -0.36  & 5     & 10    & -0.52  \\
    INS.L(58) & COTC  & -     & -     & -     & -     & -     & -     & -     & -     & -     & 13    & 20    & -0.36  \\
    SFGdor.L(59) & COTC  &       &       &       & 7     & 11    & -0.35  & -     & -     & -     & 19    & 25    & -0.23  \\
    INS.R(60) & COTC  & 11    & 15    & -0.26  & -     & -     & -     & -     & -     & -     & -     & -     & - \\
    SMG.R(62) & AUN   & -     & -     & -     & -     & -     & -     & 31    & 41    & -0.25  & 17    & 25    & -0.34  \\
    STG.R(63) & AUN   & -     & -     & -     & 28    & 39    & -0.27  & -     & -     & -     & -     & -     & - \\
    SMG.L(68) & AUN   & -     & -     & -     & 20    & 14    & 0.28  & -     & -     & -     & -     & -     & - \\
    SMG.L(69) & AUN   & -     & -     & -     & -     & -     & -     & 30    & 19    & 0.38  & 19    & 6     & 0.67  \\
    SMG.R(72) & AUN   & -     & -     & -     & -     & -     & -     & 29    & 20    & 0.32  & 21    & 12    & 0.44  \\
    PCUN.R(89) & DMN   & 23    & 18    & 0.22  & -     & -     & -     & 26    & 33    & -0.21  & -     & -     & - \\
    PCG.R(92) & DMN   & -     & -     & -     & -     & -     & -     & 15    & 11    & 0.26  & -     & -     & - \\
    CUN.R(93) & DMN   & 10    & 8     & 0.27  & -     & -     & -     & -     & -     & -     & -     & -     & - \\
    PCG.R(95) & DMN   & 11    & 15    & -0.26  & -     & -     & -     & -     & -     & -     & -     & -     & - \\
    SFGmed.R(105) & DMN   & -     & -     & -     & -     & -     & -     & 18    & 14    & 0.25  & -     & -     & - \\
    ORBsupmed.L(109) & DMN   & 30    & 20    & 0.33  & -     & -     & -     & 32    & 24    & 0.26  & -     & -     & - \\
    ACG.L(113) & DMN   & 15    & 5     & 0.67  & -     & -     & -     & -     & -     & -     & -     & -     & - \\
    PCUN.R(136) & MR    & 10    & 14    & -0.30  & -     & -     & -     & -     & -     & -     & -     & -     & - \\
    LING.R(150) & VIN   & -     & -     & -     & -     & -     & -     & -     & -     & -     & 31    & 49    & -0.36  \\
    SOG.L(155) & VIN   & -     & -     & -     & -     & -     & -     & 12    & 20    & -0.39  & 13    & 27    & -0.51  \\
    CUN.R(156) & VIN   & -     & -     & -     & -     & -     & -     & 23    & 31    & -0.28  & 22    & 32    & -0.30  \\
    SOG.R(162) & VIN   & -     & -     & -     & -     & -     & -     & -     & -     & -     & 22    & 31    & -0.29  \\
    CUN.R(163) & VIN   & 28    & 21    & 0.24  & -     & -     & -     & -     & -     & -     & -     & -     & - \\
    IPL.L(195) & FPTC  & 18    & 14    & 0.24  & -     & -     & -     & -     & -     & -     & -     & -     & - \\
    ANG.R(199) & FPTC  & 28    & 39    & -0.28  & -     & -     & -     & -     & -     & -     & -     & -     & - \\
    PCUN.R(203) & SN   & 14    & 21    & -0.32  & -     & -     & -     & -     & -     & -     & 43    & 33    & 0.23  \\
    SMG.R(204) & SN   & 5     & 2     & 0.54  & -     & -     & -     & -     & -     & -     & -     & -     & - \\
    INS.R(209) & SN   & -     & -     & -     & -     & -     & -     & 7     & 4     & 0.51  & 20    & 15    & 0.25  \\
    SMA.L(213) & SN   & 6     & 8     & -0.33  & 12    & 15    & -0.20  & -     & -     & -     & -     & -     & - \\
    SFGdor.R(216) & SN   & 2     & 5     & -0.53  & 16    & 24    & -0.31  & -     & -     & -     & -     & -     & - \\
    SPG.R(251) & DAN   & 32    & 40    & -0.21  & -     & -     & -     & -     & -     & -     & -     & -     & - \\
    IOG.L(252) & DAN   & -     & -     & -     & 48    & 35    & 0.26  & -     & -     & -     & -     & -     & - \\
    IPL.L(259) & DAN   & 33    & 26    & 0.23  & -     & -     & -     & -     & -     & -     & -     & -     & - \\
    \bottomrule
    \end{tabular}%
  \label{tab:S1}%
\end{table}%

\end{document}